\documentclass[12pt,preprint]{aastex}

\tighten

\newcommand{\asec}{\hbox to 1pt{}\rlap{$^{\prime\prime}$}.\hbox to 2pt{}}
\newcommand{\amin}{\hbox to 1pt{}\rlap{$^{\prime}$}.\hbox to 2pt{}}

\shortauthors{Lauer et al.}
\shorttitle{Biases in Evolution of Black Hole Mass Relationships}

\begin{document}

\title{Selection Bias in Observing the Cosmological Evolution of the
$M_\bullet-\sigma$ and $M_\bullet-L$ Relationships}

\author{Tod R. Lauer}
\affil{National Optical Astronomy Observatory\footnote{The National Optical
Astronomy Observatory is operated by AURA, Inc., under cooperative agreement
with the National Science Foundation.},       
P.O. Box 26732, Tucson, AZ 85726}

\author{Scott Tremaine}
\affil{Institute for Advanced Study, Einstein Drive, Princeton, NJ 08540}

\author{Douglas Richstone}
\affil{Department of Astronomy, University of Michigan, Ann Arbor, MI 48109}

\author{S. M. Faber}
\affil{UCO/Lick Observatory, Board of Studies in Astronomy and
Astrophysics, University of California, Santa Cruz, CA 95064}

\vfill


\begin{abstract}

Programs to observe evolution in the $M_\bullet-\sigma$ or
$M_\bullet-L$ relations typically compare black-hole masses, $M_\bullet,$ in
high-redshift galaxies selected by nuclear activity to $M_\bullet$ in local
galaxies selected by luminosity $L,$ or stellar velocity dispersion $\sigma.$
Because AGN luminosity is likely to depend on
$M_\bullet$, selection effects are different for high-redshift and
local samples, potentially producing a false signal of evolution.
This bias arises because cosmic scatter
in the $M_\bullet-\sigma$ and $M_\bullet-L$ relations means that the
mean $\log_{10}L$ or $\log_{10}\sigma$ among galaxies that host a black hole of given $M_\bullet$, may be substantially different than the
$\log_{10}L$ or $\log_{10}\sigma$ obtained from inverting the $M_\bullet-L$
or $M_\bullet-\sigma$ relations for the same nominal $M_\bullet.$
The bias is particularly strong at high $M_\bullet,$ where the
luminosity and dispersion functions of galaxies are falling rapidly.  The
most massive black holes occur more often as rare outliers in
galaxies of modest mass than
in the even rarer high-mass galaxies, which would otherwise be the sole
location of such black holes in the absence of cosmic scatter.
Because of this bias, $M_\bullet$ will typically
appear to be too large in the distant sample for a given $L$ or $\sigma.$ For
the largest black holes and the largest plausible cosmic scatter,
the bias can reach a factor of 3 in $M_\bullet$ for the
$M_\bullet-\sigma$ relation and a factor of 9 for the $M_\bullet-L$ relation.
Unfortunately, the actual cosmic scatter is not known well enough to correct
for the bias.  Measuring evolution of the $M_\bullet$ and galaxy property
relations requires object selection to be precisely defined and exactly
the same at all redshifts.

\end{abstract}


\section{Observing Evolution in the Relationships Between
Black Hole Mass and Galaxy Properties} 

The discovery that most elliptical galaxies and spiral bulges host a
black hole at their centers, plus the tight relations observed between
black-hole mass $M_\bullet$ and galaxy luminosity $L$ or stellar
velocity dispersion $\sigma$ \citep{d89, k93, kr, mag, fm, g00,
tr02, hr}, suggest that the formation and growth of central black
holes is deeply intertwined with that of their host galaxies.  Recent
theoretical work (e.g., \citealt{hop06}) supports this view, and also
predicts how the relations between the properties of black holes
and their host galaxies have changed over time.  Direct observation of
the evolution of the $M_\bullet-\sigma$ and $M_\bullet-L$
relations over cosmological time would offer unique insight into
galaxy and black-hole formation.

There have been many attempts to measure the evolution of the
$M_\bullet$ relations; we cite a partial list of these:

\begin{itemize}

\item \citet{sh03} examine a sample of quasars at redshifts up to 3.3. The
  luminosity $L$ and velocity dispersion $\sigma$ of the host galaxy in such
  objects cannot be measured reliably because the light from the galaxy is
  overwhelmed by the quasar flux; instead, they use the width of the narrow [O
  III] emission line as a surrogate for $\sigma$. They estimate the black-hole
  mass using the ``photoionization'' method, which is based on an empirically
  calibrated relation involving the continuum luminosity of the active
  galactic nucleus (AGN) and the width of the H$\beta$ or other broad emission
  lines. The theoretical assumptions that underlie this relation are that the
  bulk velocities of the emitting clouds are determined by their orbital
  motion in the gravitational field of the black hole, and that the emitting
  region is photoionized by an ultraviolet continuum spectrum of
  fixed shape. They find an $M_\bullet-\sigma$ relation that is consistent
  with the local one, suggesting that this relation is independent of
  redshift.

\item Using similar methods on a larger but lower-redshift quasar sample from
  the Sloan Digital Sky Survey, \citet{sal07} find that galaxies of a given
  dispersion at $z\simeq 1$ have black-hole masses that are larger by
  $\Delta\log_{10}M_\bullet\sim 0.2$ than at $z=0$.

\item \citet{tr04} and \citet{woo06} measured the velocity dispersion
  of 14 Seyfert I galaxies
  at redshift $z\simeq0.36$; they used stellar absorption lines, which should
  provide a more direct measure of the dispersion than emission lines. They
  measured the black-hole mass using the photoionization method. They found
  that galaxies of a given dispersion at $z\simeq0.36$ host black holes having
  $\Delta\log_{10}M_\bullet=0.62\pm0.10\pm0.25$.

\item \citet{peng} measured the bulge luminosities of 11 quasar hosts in the
  redshift range $1.7<z<2.7$ using the Hubble Space Telescope (HST), and estimated
  their black-hole masses using the photoionization method. They conclude that
  the $M_\bullet-L$ relation at $z\sim 2$ is close to the relation at
  $z=0$; $\Delta\log_{10}M_\bullet\simeq -0.1$. This result is remarkable
  since black holes can only grow with time, while elliptical galaxies fade
  with time as their stars die: even if the black holes do not grow at all,
  passive stellar evolution models would predict $\Delta\log_{10}M_\bullet$
  between $-0.4$ and $-0.8$ at $z\simeq 2$. If their result is correct, then
  either black holes must be ejected from the galaxy centers and replaced with
  smaller ones, or the galaxy luminosity must grow substantially through
  mergers.

\end{itemize}

A common thread among all these investigations and others is that the
high-redshift sample is selected by some measure of AGN visibility.
Black-hole masses and galaxy properties are then derived and an
$M_\bullet-\sigma$ or $M_\bullet-L$ relation is fitted to the data.  Since the
low-redshift relations are linear in $\log_{10}M_\bullet$, $\log_{10}L$, and
$\log_{10}\sigma$ (eqs. \ref{eqn:msig} or \ref{eqn:ml_hr} below), the
high-redshift data are often analyzed by assuming that the slope of the
relation is the same as at low redshift and estimating the offset in the
intercept or zero-point, which may be expressed as $\Delta\log_{10}M_\bullet$
at fixed $L$ or $\sigma$. Any such offset is then interpreted as evidence that
the ratio of $M_\bullet$ to $\sigma$ or $L$ has changed over time.  Of course,
the offset can be (and sometimes is) equally well expressed as
$\Delta\log_{10}L$ or $\Delta\log_{10}\sigma$ at fixed black-hole mass.

Proof of evolution in the $M_\bullet$ relations requires demonstrating that
the high- and low-redshift galaxy samples were assembled with no selection
effects that would bias the relation between the typical galaxy properties and
black-hole masses being measured---or, at least, that the selection effects in
the high- and low-redshift samples were the same. This may be a more difficult
task than has commonly been assumed.  The local black-hole sample has mostly
been drawn from ``normal'' galaxies with quiescent or low levels of nuclear
activity.  Despite the proximity of these galaxies, detecting and weighing
their central black holes requires exquisite spatial resolution, very high
signal-to-noise observations, and elaborate modeling of the observations.
Investigating the properties of black holes in nearby galaxies remains a
frontier problem; to date, dynamically determined black-hole masses are
available for only slightly more than three dozen galaxies. Since there was
little evidence of a black hole in most of these galaxies before HST
observations were taken (except sometimes for a modest rise in the velocity
dispersion in spectra taken at ground-based resolution), and since black holes
are found in almost all nearby galaxies that have been examined carefully with
HST, the black holes in local quiescent galaxies are selected mainly on the
basis of galaxy properties such as $L$ or $\sigma$.

The techniques used to measure black-hole masses in quiescent galaxies at low
redshift cannot be used at high redshift, both because the radius of influence
of the black hole cannot be resolved beyond a few tens of Mpc, and because the
galaxies have substantially lower surface brightness, by the factor
$(1+z)^4$. Black holes at high redshift are identified instead by their
association with AGN.  The observer thus {\it first} locates a black hole that
is accreting matter and weighs it by the properties of the AGN emission lines
(a difficult task, but one that we shall not examine in detail here), and then
{\it secondly} attempts the (still difficult) task of measuring the properties
of the host galaxy. The existence and properties of the AGN depend both on
the properties of the black hole (mass, spin, orientation) and on the properties of the
galaxy (mass inflow rate, orientation, etc.). In short, selection is done at
low redshift by galaxy properties, and at high redshift by a combination of
black-hole and galaxy properties that depends on the method used.

Galaxy samples obtained with different selection techniques will generally
satisfy different $M_\bullet$ relations. This effect, analogous to the
\citet{malm} bias that is familiar in studies of Galactic structure, arises
because of the cosmic scatter in the $M_\bullet$ relations---there is not (so
far as we know) an exact 1--1 relation between black-hole mass and any single
measurable property of galaxies such as $L$ or $\sigma$.  In this paper we
argue that determinations of the redshift evolution of the $M_\bullet$
relations may be strongly biased by selection effects unless the same sample
selection techniques are used at all redshifts. Correcting for the selection
bias introduced by the use of different sample selection criteria at different
redshifts is extremely difficult: to make accurate corrections it is necessary
to know both how the selection depends on the galaxy and black-hole properties
and the cosmic scatter in the $M_\bullet$ relations. At present we have only
crude upper limits to the latter quantity. We shall show that even if the cosmic
scatter were known, the bias in the samples described above usually can only
be corrected with additional information, such as how galaxy properties and
black-hole mass determine the probability distribution of AGN luminosity.

The selection bias can be especially large at large $M_\bullet$, as the
following argument shows. Consider the $M_\bullet-L$ relation.  At high $L$
the number density of galaxies falls off rapidly, as illustrated by the
steep cutoff in the \citet{Schechter} luminosity function.  Cosmic scatter in
the $M_\bullet-L$ relation implies that there is a distribution of black-hole
masses at a given $L$. The rare high-mass black holes can arise from either
the peak of this distribution in the rare galaxies with large $L$, or from the
high-mass tail of the distribution in the more numerous galaxies of modest
$L$. If the number density of galaxies is falling off rapidly with $L$, the
contribution from galaxies with modest $L$ may actually overwhelm the
population of black holes of similar mass associated with galaxies of higher
$L$. This problem was first explicitly identified by \citet{fi06} in the
context of the correlation between black-hole mass and dark-matter halo mass,
and by \citet{sal07} in the context of the $M_\bullet-\sigma$ relation, but
apparently has not been appreciated by most observers studying the evolution
of the $M_\bullet$ relations.

  In this paper we present a more general exposition of the biases incurred
when studying evolution of the $M_\bullet$ relations by comparing samples
obtained by different selection criteria and at different redshifts.  We start
by considering the extraction of samples from hypothetical joint distributions
of $M_\bullet$ and $L$ or $\sigma.$ We then consider the selection bias that
occurs if the high-redshift samples are selected by AGN flux or luminosity.
We discuss the prospects of correcting for selection bias, and argue that
selection bias may place fundamental limits on the determination of
the evolution of the $M_\bullet$ relations. We conclude with a brief review of
attempts to measure the evolution of the $M_\bullet$ relations and how they
may have been affected by selection bias. 

\section{The Joint Distribution of Black-Hole Mass and Galaxy Properties}

Understanding object selection bias requires knowledge of the joint
probability distribution of black-hole mass $M_\bullet$ and a galaxy property
$s,$ which for the present discussion is either $\log_{10}L$, where $L$ is the
rest-frame $V$-band galaxy luminosity, or $\log_{10}\sigma$, where $\sigma$ is
the line-of-sight velocity dispersion in the main body of the galaxy (the
precise definition of $L$ or $\sigma$ does not concern us, so long as it is
defined consistently in all samples).  The true form of the joint distribution
unfortunately is unknown; however, we can construct a hypothetical form of
this distribution that accurately represents the present state of the
observations, and in any case suffices to show how sample selection bias can
occur.

We write the probability of finding a galaxy in the
interval $(\mu,\mu+d\mu)$ and $(s,s+ds)$ as $\nu(\mu,s)\,d\mu\,ds$
where $\mu=\log_{10}(M_\bullet)$. From the definition of conditional
probability this can be rewritten as 
\begin{equation} 
  \nu(\mu,s)=\nu(\mu|s)g(s),
\end{equation} 
where $g(s)\,ds$ is the probability that a randomly chosen galaxy
in a given volume lies in the interval $(s,s+ds)$ and
$\nu(\mu|s)\,d\mu$ is the probability that the black-hole mass lies in
the range $(\mu,\mu+d\mu)$ given that the galaxy property is $s$. The
function $g(s)$ is then given by either the volume-limited luminosity
function---more properly, the luminosity function of early-type galaxy
components (ellipticals and spiral bulges), since these are the
components that correlate with black-hole mass---or the
velocity-dispersion distribution of early-type components. Without
loss of generality, we may write $\nu(\mu,s)=h[\mu-f(s),s]$ where
$\int h(x,y)dx=1$ and $\int x h(x,s)dx=0$, so $f(s)$ is the mean value
of $\mu$ at a given value of the galaxy property $s$---that is, $f(s)$
is either the $M_\bullet-L$ or $M_\bullet-\sigma$ relation. The
limited observational data on black-hole masses in local galaxies is
consistent with the hypothesis that $h(x,s)$ is independent of $s;$
adopting this hypothesis for simplicity we have
\begin{equation}
	\nu(\mu,s)=h\left[\mu-f(s)\right]g(s).
\label{eqn:joint}
\end{equation}
The intrinsic variance or ``cosmic scatter'' in black-hole mass at a given
value of galaxy property $s$ is $\sigma_\mu^2=\int x^2h(x,s)dx.$ 
Observational errors may make $h(x,s)$ appear to be yet broader, but
for this discussion, we assume that observational errors are negligible;
the selection effects that we are concerned with are solely due to the fact
that there is an intrinsic and irreducible range of $M_\bullet$ at any $s.$

It should be stressed that although the functional form
(\ref{eqn:joint}) is consistent with the available data, it is far
from unique. As a foil, we may consider the form
\begin{equation}
	\nu(\mu,s)=p\left[s-c(\mu)\right]\Phi_\bullet(\mu).
\label{eqn:jointa}
\end{equation}
A physical model that motivates (\ref{eqn:joint}) is one in which the galaxy
property $s$ determines the black-hole mass $\mu$ with a cosmic scatter
described by the function $h$, while (\ref{eqn:jointa}) is motivated by models
in which the black-hole mass $\mu$ determines the galaxy property $s$, with
cosmic scatter described by the function $p$. If $\int p(y)dy=1$, then
$\Phi_\bullet(\mu)d\mu$ is the number of black holes per unit volume with log
mass in the range $(\mu,\mu+d\mu).$ We do not know which of (\ref{eqn:joint})
or (\ref{eqn:jointa}) is correct (possibly neither, or both); our motivation
for choosing the former is that it provides a simple representation of what
local observations of black holes in inactive galaxies measure.

\begin{figure}
\plotone{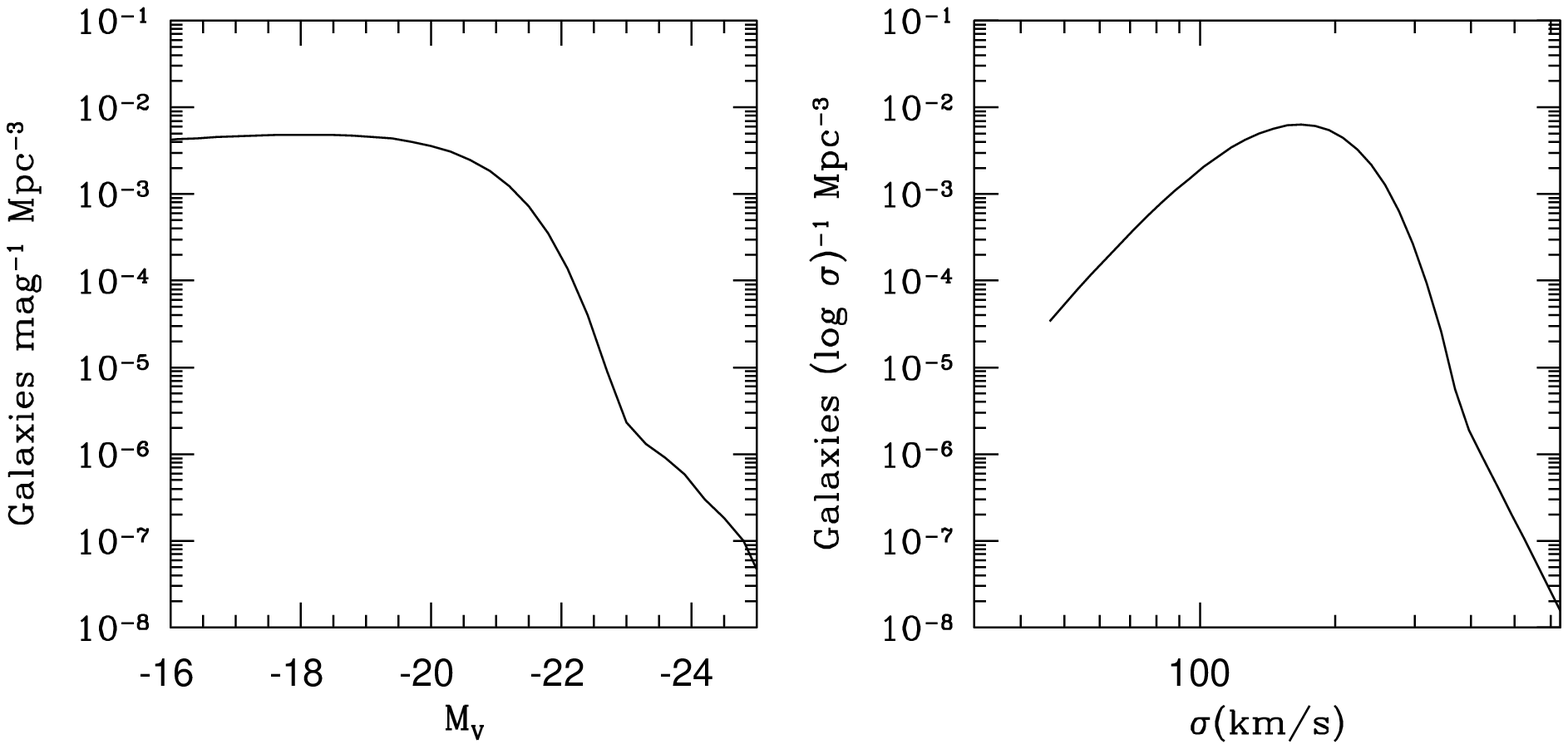}
\caption{The number density of galaxies as a function of luminosity (left) and
  velocity dispersion (right), assuming $H_0=70~{\rm km~s^{-1}\,Mpc^{-1}}$.
  The luminosity function is based on data from the Sloan Digital Sky Survey
  \citep{blanton}, augmented at large luminosities by the \citet{pl}
  sample of brightest cluster galaxies. The velocity-dispersion function is
  based on SDSS data \citep{sheth}, augmented at large dispersions by the
  sample identified by \citet{bern3}.  See \citet{l07} for details.}
\label{fig:g}
\end{figure}

Of the three functions that contribute to $\nu(\mu,s)$, $g(s)$ is
probably the best determined.  We discuss the galaxy-property
functions in detail in \citet{l07}, but summarize them briefly here.
For the galaxy luminosity function we use the \citet{blanton} Sloan
Digital Sky Survey (SDSS) luminosity function, transformed to the
$V$-band and redshift $z=0.1$, augmented with the \cite{pl} luminosity
function of brightest cluster galaxies, which appear to be
undercounted in the SDSS.  We note that the Blanton et al.\ function
refers to total rather than bulge luminosity in S0 and spiral
galaxies; however, this difference is less important at the bright end
of the luminosity function, where the selection effects are strongest.
For the velocity-dispersion function we use the \citet{sheth} SDSS
results, augmented at the high-$\sigma$ end as prescribed by
\citet{bern3} to account for an artificial high-$\sigma$ cut-off in
Sheth et al.  Both functions are shown in Figure \ref{fig:g}.

The function $f(s)$ should be the least-squares fit of $\log_{10}M_\bullet$ to
$\log_{10}L$ or $\log_{10}\sigma$ in a sample selected by galaxy properties.
The $M_\bullet-\sigma$ relation
$f(s)$ is based on the galaxy sample from \citet{tr02}
augmented by a few galaxies with more recent $M_\bullet$ determinations
(see \citealt{l07}). In contrast to the treatment in \citet{tr02}, which
treated $M_\bullet$ and $\sigma$ symmetrically, the appropriate treatment for our purposes is a least-squares fit
of $\log_{10}M_\bullet$ on $\log_{10}\sigma$, which gives:
\begin{equation}
       \log_{10} (M_\bullet/M_\odot)=(4.13\pm0.32)\log_{10}(\sigma/200
       {\rm\ km\ s^{-1}})+8.29\pm0.07,
\label{eqn:msig}
\end{equation}
for $H_0=70~{\rm km~s^{-1}\,Mpc^{-1}}$ (which we will use throughout
this paper).  We discuss the $M_\bullet-L$ relation in detail in
\citet{l07}; in brief we use the \citet{hr} relation between
$M_\bullet$ and galaxy {\it mass}, transformed back to luminosity by
adopting the mass-to-light ratio
$M/L_V\simeq6\times10^{-0.092(M_V+22)}M_\odot/L_\odot,$ based on the
$M/L$ estimates given in \citet{g03}. A least-squares fit
of $M_\bullet$ on $M_V$ for galaxies with $M_V<-19$ gives:
\begin{equation}
    \log_{10} (M_\bullet/M_\odot)=(1.32\pm0.14)(-M_V-22)/2.5+8.67\pm0.09.
\label{eqn:ml_hr}
\end{equation}

Equations (\ref{eqn:msig}) and (\ref{eqn:ml_hr}) have the form
\begin{equation}
\label{eqn:msrel}
     f(s)=a+bs,
\end{equation}
where $a$ and $b$ are constants. \citet{l07} show that the $M_\bullet-\sigma$
and $M_\bullet-L$ relations make different predictions for $M_\bullet$ in
luminous or high-dispersion galaxies, which have lower $\sigma$ values for a
given luminosity than is implied by eliminating $M_\bullet$ from equations
(\ref{eqn:msig}) and (\ref{eqn:ml_hr}).  Thus at least one of the
$M_\bullet-\sigma$ or $M_\bullet-L$ relations must curve away from the linear
fit (\ref{eqn:msrel}) in the most luminous galaxies. We are not concerned with
this issue here, and simply show the different selection biases that can
result from using $\sigma$ or $L$ as the galaxy property that correlates with
$M_\bullet$, {\it assuming} that the linear relation (\ref{eqn:msrel}) holds
at all masses.  However, the reader should bear in mind that neither the
$M_\bullet-L$ or $M_\bullet-\sigma$ relation is well determined at high galaxy
masses, where the potential selection bias is most important.

The scatter function, $h\left[\mu-f(s)\right],$ is poorly known at best.
For this analysis, we assume that at any galaxy $L$ or $\sigma$,
$\log_{10} M_\bullet$ is described by a normal distribution about
$f(s)$, with cosmic scatter $\sigma_\mu$. Thus
\begin{equation}
   h\left[\mu-f(s)\right]\,d\mu= {d\mu\over \sqrt{2 \pi{\sigma_\mu}^2}}
   \exp\left[ -
   {[\mu - f(s)]^2 \over 2{\sigma_\mu}^2}\right].
\label{eqn:prob}
\end{equation}
Again, we emphasize that $\sigma_\mu$ does not embody any observational
errors in the determination of $M_\bullet;$ it represents the
intrinsic spread in $M_\bullet$ at any galaxy property.  \citet{novak}
conclude that only an upper limit to $\sigma_\mu$ is presently known
for either the $M_\bullet-\sigma$ or $M_\bullet-L$ relation, both because
of the small sample of reliable $M_\bullet$ determinations, and because of
uncertainties in the observational errors in $M_\bullet$, which must be
accurately measured to isolate the contribution of $\sigma_\mu$ to the
total residuals.  For the present analysis we will explore the
sensitivity of the selection biases to $\sigma_\mu$ on the assumption
that $\sigma_\mu<0.3$ for dispersion $\sigma$ and $\sigma_\mu<0.5$ for
luminosity $L$. 

Apart from the poor knowledge of $\sigma_\mu$, there are at least two other
major uncertainties in the function $h$: (i) a more general treatment would
allow for the possibility that it varies with $s$; (ii) there is little or no
justification of the assumed normal form.  As will be shown below, for the
most massive black holes or galaxies, the selection bias may depend on the
form that $h\left[\mu-f(s)\right]$ takes at several standard deviations away
from the mean, yet an observational determination of the form of $h$ in this
region would require a sample of well-determined black-hole masses several
orders of magnitude larger than is presently available.  Thus at the highest
black-hole masses, the selection bias is likely to depend sensitively on
knowledge that is not presently at hand, so it is not possible to apply
reliable corrections for this bias.

\begin{figure}
\plotone{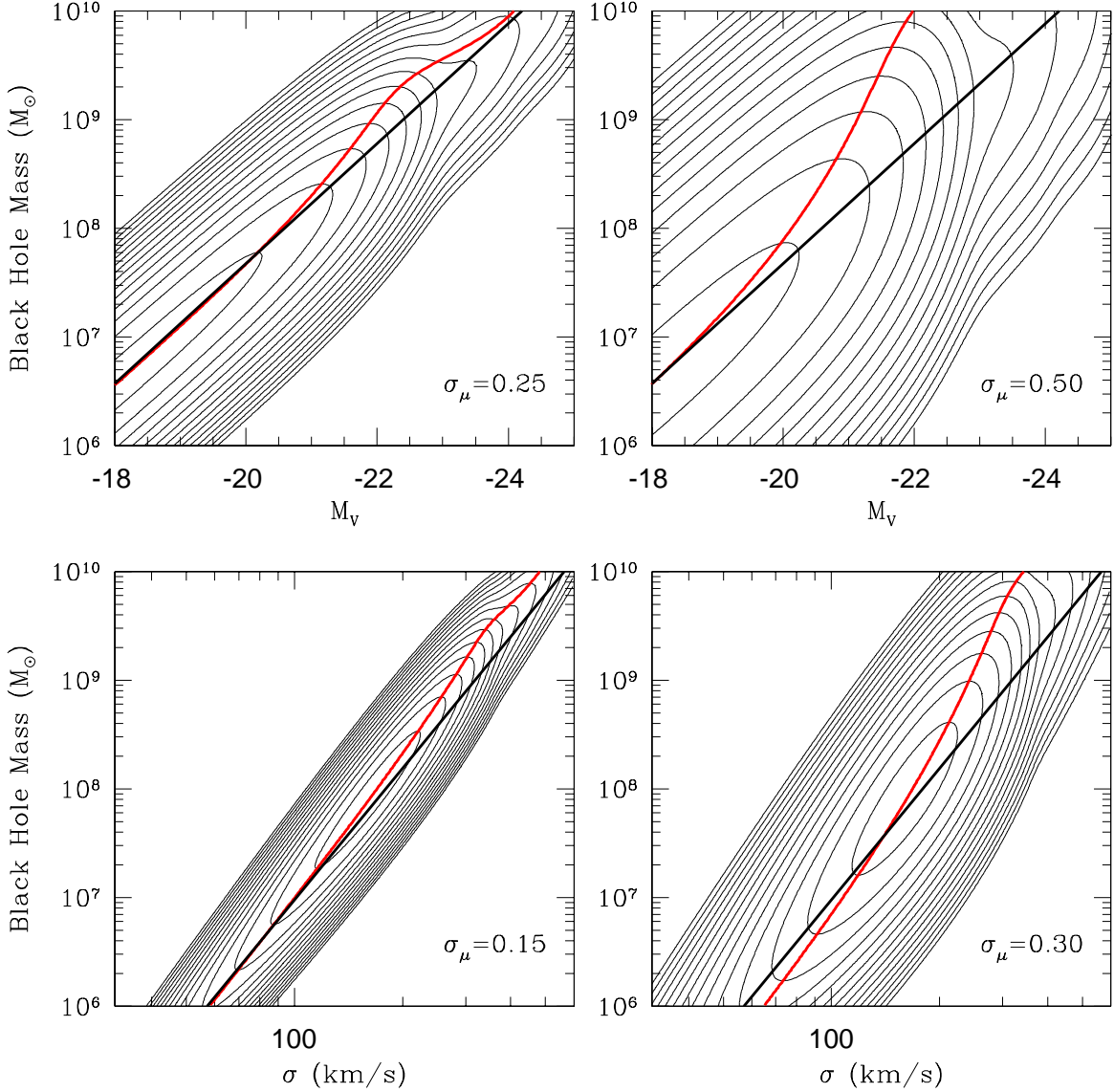}
\caption{The joint probability distribution of galaxy luminosity $L$ (top
panels) or stellar velocity dispersion $\sigma$ (bottom panels)
and black-hole mass, $M_\bullet.$
The solid, straight, black line gives the mean $M_\bullet-L$ or
$M_\bullet-\sigma$ relation (eqs. \ref{eqn:msig} and \ref{eqn:ml_hr}).  The
joint distribution is calculated by assuming that the black-hole mass has a
log-normal distribution about the $M_\bullet-L$ or $M_\bullet-\sigma$
relation, with the normalization provided by the galaxy luminosity or
velocity-dispersion functions shown in Figure \ref{fig:g}.  The adopted
dispersions in $\log_{10}M_\bullet$ are $\sigma_\mu=0.25,\ 0.50$ for the
$M_\bullet-L$ relation or $\sigma_\mu=0.15,\ 0.30$ for the $M_\bullet-\sigma$
relation.  The contours are arbitrary, but are spaced in increments of 0.5 dex
in density.  The curved red lines give the mean $M_V$ or $\log_{10}(\sigma)$
as a function of $M_\bullet.$ The displacement of this line from the
mean relations gives the bias at a given $M_\bullet.$}
\label{fig:2df}
\end{figure}

\begin{figure}
\plotone{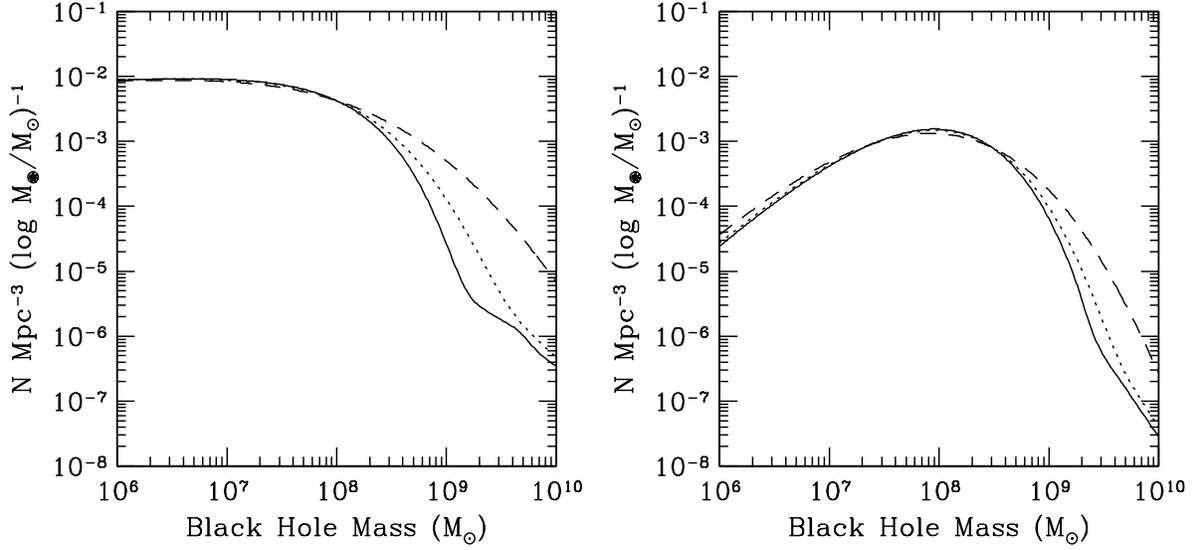}
\caption{The effects of cosmic scatter in the $M_\bullet-\sigma$ or
$M_\bullet-L$ relations on the number density of black holes as a function of
$M_\bullet$.  The left panel is based on the $M_\bullet-L$ relation of
equation (\ref{eqn:ml_hr}) combined with the luminosity function of Figure
\ref{fig:g}.  The right panel is based on the $M_\bullet-\sigma$ relation of
equation (\ref{eqn:msig}) combined with the velocity-dispersion function of
Figure \ref{fig:g}.  Solid lines show the results assuming no scatter in the
$M_\bullet$ relations, while the dashed lines show the effects of increasing
amounts of cosmic scatter $\sigma_\mu$. The dotted and dashed lines show the
effects of $\sigma_\mu=0.25,0.50$ for the $M_\bullet-L$ relationship and
$\sigma_\mu=0.15,0.30$ for the $M_\bullet-\sigma$ relationship. Cosmic scatter
greatly enhances the volume density of black holes at the high-mass end.}
\label{fig:bh_df}
\end{figure}

\begin{figure}
\includegraphics[bb=200 170 400 670]{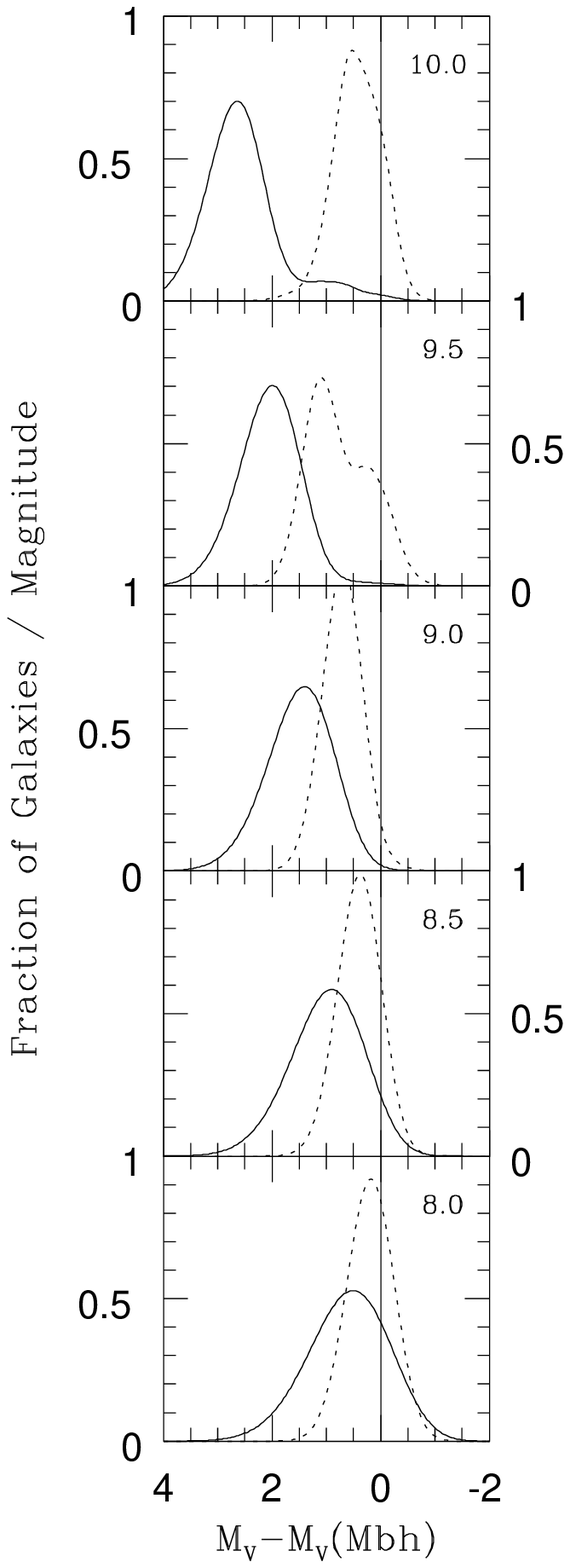}
\includegraphics[bb=200 170 400 670]{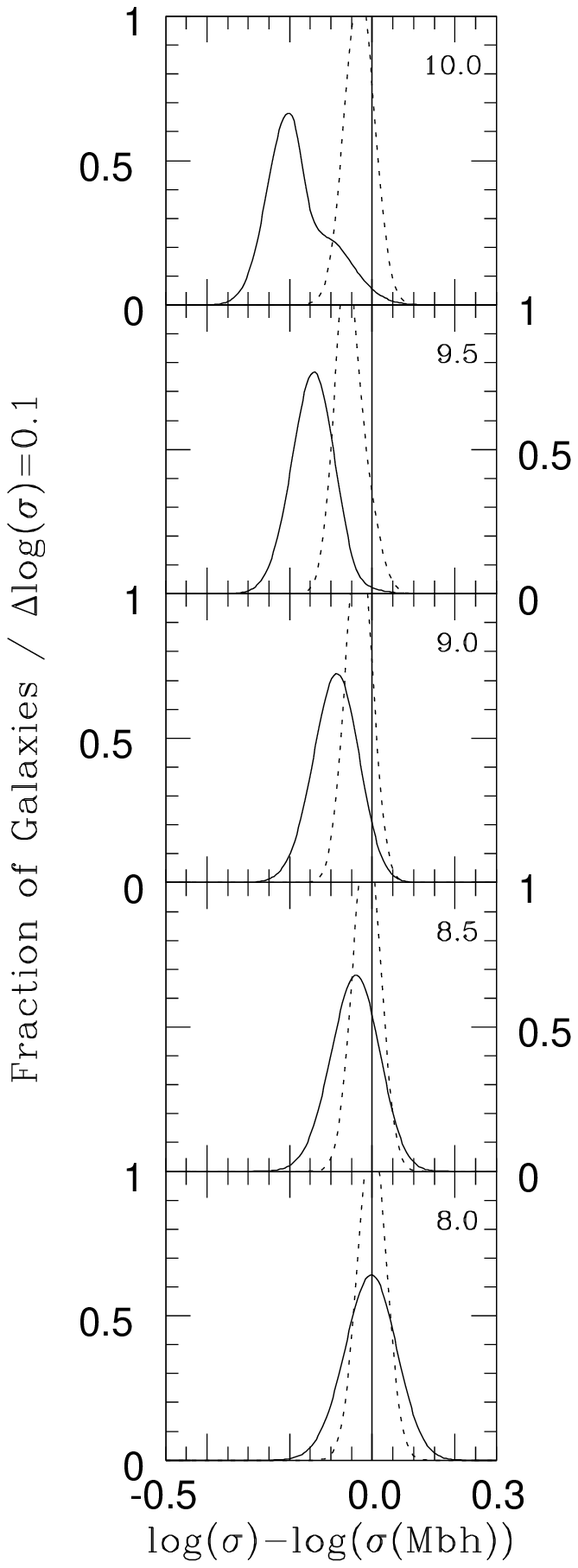}
\caption{The distribution of absolute magnitude $M_V=-2.5\log_{10}
L+\hbox{const}$ or velocity dispersion $\log_{10}\sigma$ at selected values of
$M_\bullet$ ($\log_{10}M_\bullet/M_\odot$ is given in the
upper right of each panel). These probability distributions are obtained by
taking horizontal cuts through the joint distributions shown in Figure
\ref{fig:2df}, and are appropriate when the sample selection is on black-hole
mass. Two distributions are shown with dashed and solid lines, corresponding
respectively to $\sigma_\mu=0.25,\ 0.50$ for $M_V$ on the left or
$\sigma_\mu=0.15,\ 0.30$ for $\log_{10}\sigma$ on the right. The horizontal
coordinate is the difference between either $M_V$ or $\log_{10}\sigma$ and the
nominal values $f^{-1}(s)$ determined from the $M_\bullet-L$ or
$M_\bullet-\sigma$ relations (\ref{eqn:msig}) and (\ref{eqn:ml_hr}).}
\label{fig:bh_cuts}
\end{figure}

With these caveats, in Figure \ref{fig:2df} we show estimates of $\nu(\mu,s)$
as modeled by equation (\ref{eqn:joint}) for each of the $M_\bullet-\sigma$
and $M_\bullet-L$ relations, and two assumed values of $\sigma_\mu.$
Projection of $\nu(\mu,s)$ onto the $\mu$-axis produces the mass function of
black holes,
\begin{eqnarray}
\Phi_\bullet(\mu)&=&\int \nu(\mu,s)\,ds \\
         &=&\int h\left[\mu-f(s)\right]g(s)\,ds.
\end{eqnarray}
The importance of cosmic scatter for converting the luminosity or dispersion
function $g(s)$ to a black-hole mass function has been discussed several times
\citep{yu02, yulu, tundo, l07}.  What may be less appreciated, however, is
that cosmic scatter implies that the most massive BHs are often hosted by
modest galaxies that {\it a priori} would not be expected to harbor BHs of
high mass. To illustrate this point, Figure \ref{fig:bh_df} shows
$\Phi_\bullet(\mu)$ for four different versions of $\nu(\mu,s).$ The salient
feature is that as the cosmic scatter $\sigma_\mu$ increases, the contribution
to the density of the most massive black holes from the wings of the scatter
function $h$ overwhelms the ``native'' population of massive black holes
harbored by the galaxies with the largest values of property $s$.  A second
illustration of this point occurs in the top two panels in Figure
\ref{fig:bh_cuts}, which show the probability distribution of $L$ and $\sigma$
for $M_\bullet=10^{10}M_\odot$: note that the solid curves, corresponding to
$\sigma_\mu=0.5$ for $L$ and 0.3 for $\sigma$, have a prominent shoulder to the right of the peak, the
peak arising from low-luminosity or low-dispersion galaxies, and the shoulder
from high luminosities and dispersions.

The source of the selection bias can be seen more directly in Figure
\ref{fig:2df}, which shows $\nu(\mu,s)$ for each of the $M_\bullet-\sigma$ and
$M_\bullet-L$ relations, and two assumed values of $\sigma_\mu$.  Only for
$\sigma_\mu=0$ is there an exact relation between $M_\bullet$ and $\sigma$
or $L.$ Our model for $\nu(\mu,s)$---which may not be correct, but is
consistent with the data---implies that the density contours in Figure
\ref{fig:2df} are symmetric about the mean relationship ridge-lines in the
vertical direction; in other words, the conditional probability of $\mu$ given
$s$ is symmetric about $\mu=f(s)$. On the other hand, the conditional
probability of $s$ given $\mu$,
\begin{equation}
\nu(s|\mu)={\nu(\mu,s)\over \int \nu(\mu,s)\,ds},
\label{eq:scond}
\end{equation}
is not symmetric about $s=f^{-1}(\mu)$, as seen in Figure \ref{fig:bh_cuts},
which shows the distribution of $s$ at selected values of $M_\bullet.$ As
$M_\bullet$ and $\sigma_\mu$ increase, the distribution of $s$ for a given
$M_\bullet$ moves further and further away from the value implied by the
inverse of the $M_\bullet$ relations, $f^{-1}(\mu)=(\mu-a)/b$, where $a$ and
$b$ are the coefficients given in equation (\ref{eqn:msrel}).
For $M_\bullet>10^9M_\odot,$ $f^{-1}(\mu)$ generally falls well out
in the wings of the
distribution of $s$ for both $L$ and $\sigma,$ particularly for the
larger values of $\sigma_\mu.$
The galaxies with
$s=f^{-1}(\mu)$ for the highest mass black holes are so far down in the
step cutoffs of the $L$ and $\sigma$ distribution functions that they
are completely overwhelmed by the population of galaxies of modest mass that
harbor high mass black holes as statistical outliers.

The mean value of $s$ at a given $M_\bullet$ is 
\begin{equation}
\langle s\rangle_\mu=
\int s\,\nu(s|\mu)\,ds={\int s\,\nu(\mu,s)\,ds\over\int \nu(\mu,s)\,ds}
={\int s\,g(s) h\left[\mu-f(s)\right]\,ds\over
       \int g(s) h\left[\mu-f(s)\right]\,ds}.
\label{eqn:means}
\end{equation}
This is shown as the red lines in Figure \ref{fig:2df}.  The mean of
$\log_{10}L$ or $\log_{10}\sigma$ at a given $M_\bullet$ is offset from
$f^{-1}(\mu),$ which is just the $s$ location of the $M_\bullet$ relation
ridgelines shown in the Figure.  In the presence of cosmic scatter,
the mean $s$ of a sample of galaxies that host a black hole of given $\mu$ is
different from the mean $\mu$ hosted by a sample of galaxies of given $s$.
This difference is the source of the selection bias.

\section{Illustration of Selection Biases}

Selection bias typically occurs because the galaxy samples used to probe the
$M_\bullet$ relations at cosmological distances are not selected by galaxy $L$
or $\sigma$, but by the visibility of their AGNs.  For most of the discussion
in this section we shall assume that AGN luminosity does not depend directly
on $L,$ $\sigma,$ or any other galaxy property unrelated to
$M_\bullet.$ This model is appropriate if, for example, the probability that
the AGN associated with a black hole of mass $M_\bullet$ has luminosity
$L_{\rm AGN}$ is given by
\begin{equation} 
  dp = \psi(\lambda-\mu)\,d\lambda \quad\hbox{where}\quad \int\psi(x)\,dx=1  
\label{eq:probone}
\end{equation}
and $\lambda\equiv\log_{10}L_{\rm AGN}$ and $\mu=\log_{10}M_\bullet$.  The
physical content of this assumption is that the luminosity history of an AGN
is determined by the black-hole mass and scales with the Eddington luminosity,
which is proportional to $M_\bullet$.

\subsection{Bias in a Luminosity-Limited Survey}
\label{sec:bhmass}

We first consider the bias that occurs in a sample in which (i) all of the
objects are in a narrow redshift range; (ii) the survey contains all AGNs
brighter than a given flux. In this case the probability that an AGN in the
relevant redshift range is accepted in the survey depends only on its
luminosity $L_{\rm AGN}$, and by equation (\ref{eq:probone}) this in turn depends
only on its black-hole mass $\mu$. Thus the probability distribution of a
galaxy property $s$ at a given value of $\mu$ is not biased by the selection,
and is simply $\nu(s|\mu)$ (eq.\ \ref{eq:scond}).  Some examples are shown in
Figure \ref{fig:bh_cuts}, which plots the probability distribution of
$s-f^{-1}(\mu)$ for several values of black-hole mass $M_\bullet$ and cosmic
scatter $\sigma_\mu$. Here $f^{-1}(s)$ is the inverse of the $M_\bullet-L$ or
$M_\bullet-\sigma$ relation (eq.~\ref{eqn:joint}). The mean values of the
distributions $\langle s\rangle_\mu$ are displaced to values lower than $f^{-1}(\mu)$ in
most of the examples shown.  As expected, the width and offset of the
distributions are larger for larger values of $\sigma_\mu$. Also as expected,
the offsets are generally larger for more massive black holes.  $\langle
s\rangle_\mu$ is shown as the red lines in Figure \ref{fig:2df}; the offset of
these lines from the relation $\mu=f(s)$ demonstrates the selection bias.

In this model, the selection bias can be corrected: an assumed form for the
probability distribution $\nu(\mu,s)$ such as (\ref{eqn:joint}) with
redshift-dependent parameters can be used to compute $\nu(s|\mu)$, which can
then be fitted to the distribution of galaxy property $s$ in the sample at a
given value of black-hole mass $\mu$ (separate issues, discussed in
\S\ref{sec:difficult}, are whether this model for the selection effects is
realistic and whether the assumed form of $\nu(\mu,s)$ is correct). However, most papers in the literature have not taken this
approach, so it is worthwhile to estimate the biases that might be introduced
by using simpler statistics to estimate the evolution in the $M_\bullet$
relations. 

\begin{figure}
\plotone{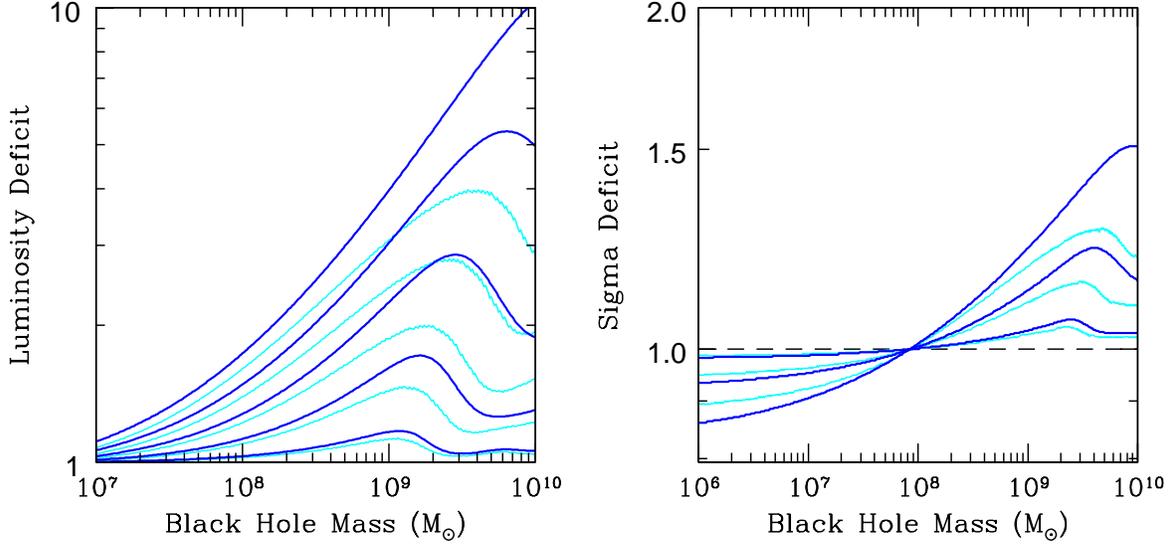}
\caption{The ratio of the nominal values $f^{-1}(s)$ determined from the
$M_\bullet-L$ or $M_\bullet-\sigma$ relations to $\langle s\rangle_\mu,$ the
mean of $\log_{10} L$ or $\log_{10} \sigma$ at a given black-hole mass. This
Figure quantifies the bias in a sample selected by $M_\bullet$
as opposed to $s.$ The blue curves  are drawn for
$\sigma_\mu=0.1,0.2,\ldots,0.5$ for $\log_{10} L$ (left panel), and
$\sigma_\mu=0.1,0.2,0.3$ for $\log_{10}\sigma$ (right panel). The cyan curves
show the biases that result if the assumed normal distributions in
$\log_{10}M_\bullet$ (eq.\ \ref{eqn:prob}) are truncated at $\pm2\sigma_\mu.$
At high $M_\bullet$ the bias depends not only on the amplitude of the
cosmic scatter, but also on the form of the wings of the scatter function.}
\label{fig:bh_bias}
\end{figure}

Figure \ref{fig:bh_bias} shows the object selection bias $\Delta s=f^{-1}(\mu)
-\langle s\rangle_\mu$ as a function of $M_\bullet$ and $\sigma_\mu$.
In this example we have expressed the bias in terms of galaxy properties, since
we have selected by $M_\bullet;$ however, it may also be represented as
$\Delta \log_{10}M_\bullet=b\Delta s,$ where $b$ is the slope given in
equation (\ref{eqn:msrel}). Note that the bias is not always a monotonic
function of black-hole mass, nor does it always have the same sign. In this
simple model, the bias at black-hole masses of $10^9M_\odot$ is
$\Delta\log_{10}M_\bullet=0.4$ in the $M_\bullet-\sigma$ relation for
$\sigma_\mu=0.3$, and $\Delta\log_{10}M_\bullet=0.7$ in the $M_\bullet-L$
relation for $\sigma_\mu=0.5$---and even larger at larger black-hole
masses. The bias is smaller at lower $L$ or $\sigma$, but we caution that the
near-zero bias in the $M_\bullet-L$ relation at small $M_\bullet$ is an
artifact of our assumption that the luminosity function has the
\citet{Schechter} form
at low luminosities (see below), and this may not be true for the early-type
galaxy components that are believed to host the black holes. 

Figure \ref{fig:bh_bias} also shows that a significant portion
of the bias comes from galaxies with $|s-f^{-1}(\mu)|>2\sigma_\mu$
for the larger values of $\sigma_\mu.$
Indeed for $M_\bullet\approx5\times10^9M_\odot,$ nearly half of the
bias comes from such galaxies for all but the smallest $\sigma_\mu$ shown.
Thus, not only is the bias sensitive to
$\sigma_\mu,$ it also depends on the shape of the wings of
the error distribution.  If, as is likely, the wings are more extended than in
our assumed log-normal distribution, the bias will be even larger.

Many of the features in these plots can be understood analytically. From the
definition of $\langle s\rangle_\mu$ (equation \ref{eqn:means}) and equation
(\ref{eqn:msrel}), we can write
\begin{equation}
a+b\langle s\rangle_\mu=\mu - {\int x\,h(x)g[b^{-1}(\mu-a-x)]\,dx\over
\int h(x)g[b^{-1}(\mu-a-x)]\,dx}.
\label{eq:meanss}
\end{equation}
If the cosmic scatter $\sigma_\mu^2=\int dx\,x^2h(x)$ is not too
large, we can evaluate this by expanding $g(s)$ in a Taylor series,
\begin{equation}
a+b\langle s\rangle_\mu=\mu + \sigma_\mu^2\left[d\ln g(s)\over 
ds\right]_{s=(\mu-a)/b}+\hbox{O}(\sigma_\mu^4).
\label{eq:meansss}
\end{equation}
This result shows that (i) the bias is proportional to $\sigma_\mu^2$
if $\sigma_\mu$ is not too large; (ii) the bias in $\langle s\rangle_\mu$
is positive if $g(s)$ declines with $s$ and proportional to $d\ln
g(s)/ds$---this is why the bias is large and negative for the most
luminous or high-dispersion galaxies; (iii) the bias is near zero if
$g(s)$ is constant, which corresponds to a luminosity function
$dn\propto dL/L$ when $s=\log_{10}L$. The bias in the left panel of
Figure \ref{fig:bh_bias} is small for low-luminosity galaxies because
the assumed luminosity function is close to this form.

A different statistical method, which more closely parallels the approach
used in most observational papers, is simply to estimate the average value of
the difference between $\mu$ and the prediction of the $\mu-s$ relation
(\ref{eqn:msrel}) for all the galaxies in the sample, 
\begin{equation}
\Delta\log_{10}M_\bullet\equiv
\langle\mu-f(s)\rangle=\langle\mu\rangle-a-b\langle s\rangle,
\end{equation}

Let us assume that the survey flux limit at the given redshift corresponds to
a luminosity $L_0$ with $\lambda_0=\log_{10}L_0$. Then
\begin{eqnarray}
\Delta\log_{10}M_\bullet&=& {\int_{\lambda_0}^\infty d\lambda 
\int g(s)\,ds\int (\mu-a-bs)h(\mu-a-bs)\psi(\lambda-\mu)\,d\mu\over 
\int_{\lambda_0}^\infty d\lambda 
\int g(s)\,ds\int h(\mu-a-bs)\psi(\lambda-\mu)\,d\mu} \\
&=& {\int_{\lambda_0}^\infty d\lambda 
\int g(s)\,ds\int xh(x)\psi(\lambda-x-a-bs)\,dx\over 
\int_{\lambda_0}^\infty d\lambda 
\int g(s)\,ds\int h(x)\psi(\lambda-x-a-bs)\,dx} \\
&=& {\int_{\lambda_0}^\infty d\lambda \int xh(x)\phi(\lambda-x)\,dx\over 
\int_{\lambda_0}^\infty d\lambda \int h(x)\phi(\lambda-x)\,dx},
\label{eq:xxxx}
\end{eqnarray}
where $x\equiv \mu-a-bs$, and
\begin{equation}
	\phi(y)=\int g(s)\psi(y-a-bs)\,ds.
\label{eq:defphi}
\end{equation}

The luminosity function (number per unit volume) of AGNs is
$dn=\Psi(\lambda)d\lambda$, where
\begin{equation}
\label{eq:phidef}
	\Psi(\lambda)=\int g(s)\,ds\int
	h(\mu-a-bs)\psi(\lambda-\mu)\,d\mu=\int h(x)\phi(\lambda-x)\,dx,
\label{eq:convolve}
\end{equation}
which is just the inner integral of the denominator of (\ref{eq:xxxx}). To
evaluate the inner integral in the numerator of equation (\ref{eq:xxxx}), note
that the functional form of $h(x)$ (eq.~\ref{eqn:prob}) implies that
$h'(x)=-xh(x)/\sigma_\mu^2$, so
\begin{equation}
\int x\,h(x)\phi(\lambda-x)\,dx=-\sigma_\mu^2\int h'(x)\phi(\lambda-x)\,dx.
\end{equation}
Integrating the right side of the equation by parts, and noting
that $\lim_{y\rightarrow\infty}\phi(y)=0,$ gives
\begin{eqnarray}
\int x\,h(x)\phi(\lambda-x)\,dx&=&\sigma_\mu^2\int h(x){\partial\over\partial x}\phi(\lambda-x)\,dx\\
   &=&-\sigma_\mu^2\int h(x){\partial\over\partial\lambda}\phi(\lambda-x)\,dx\\
   &=&-\sigma_\mu^2{d\over d\lambda}\int h(x)\phi(\lambda-x)\,dx.
\label{eq:parts}
\end{eqnarray}
Thus
\begin{equation}
\Delta\log_{10}M_\bullet ={\sigma_\mu^2\Psi(\lambda_0)
\over\int_{\lambda_0}^\infty \Psi(\lambda\,)d\lambda}. 
\label{eq:biasone}
\end{equation}
Remarkably, the result depends only on the directly observable luminosity
function $\Psi(\lambda)$ for the type of AGN targeted in the survey, and is
independent of assumptions about the luminosity history $\psi(\lambda-\mu)$ or
the distribution of host galaxy properties $g(s)$. Also, in contrast to the
analogous result (\ref{eq:meansss}), equation (\ref{eq:biasone}) is not just
the first term in a Taylor series in $\sigma_\mu^2$ but valid for all values
of $\sigma_\mu^2$, no matter how large.

\begin{figure}
\plotone{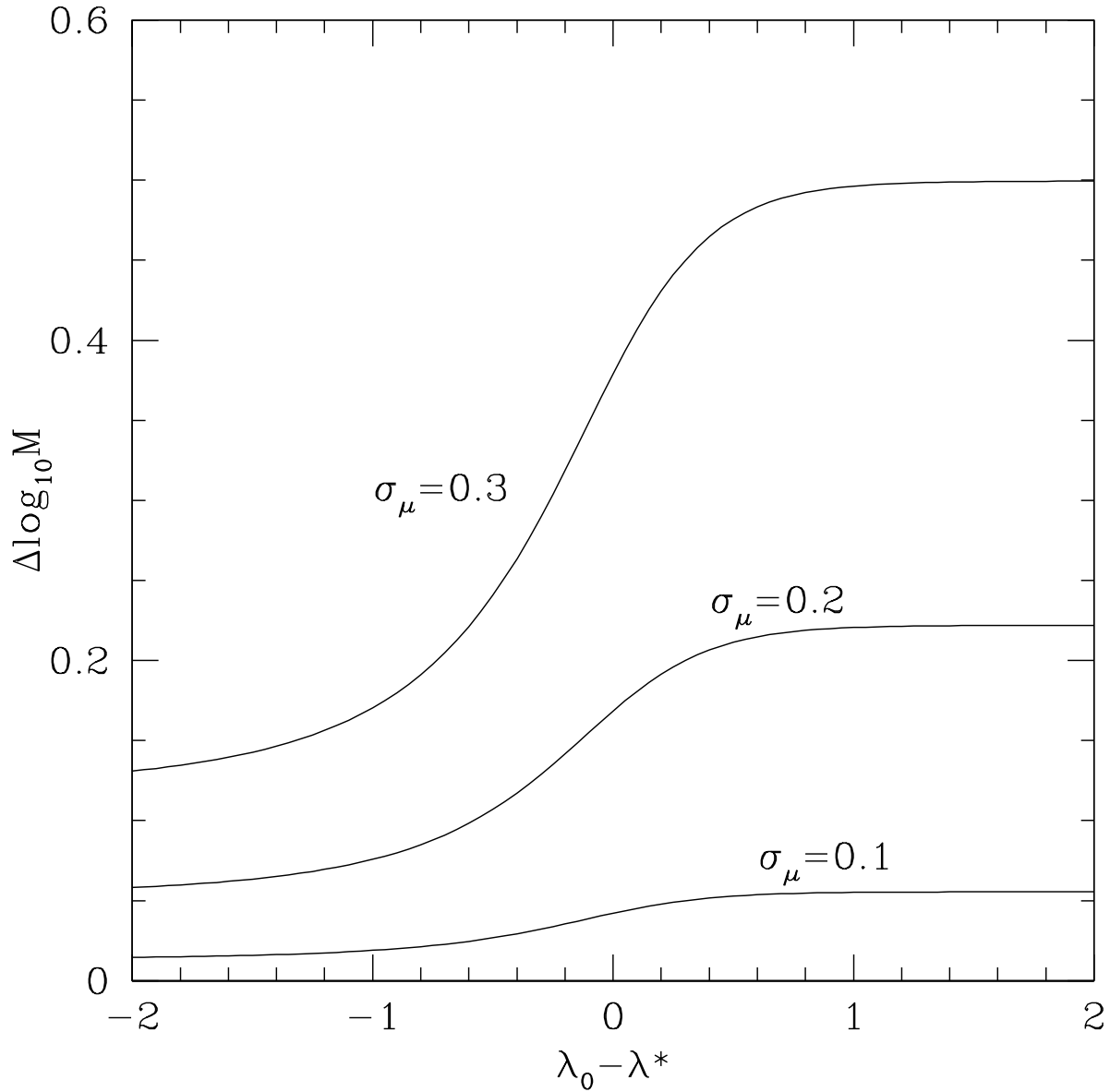}
\caption{The bias in $\log_{10}M_\bullet$ from the $M_\bullet-\sigma$ relation
for inactive galaxies in a luminosity-limited survey of AGN. The horizontal axis is
$\lambda_0-\lambda^*=\log_{10}(L_0/L^\ast)$ where $L_0$ is the limiting
luminosity at the survey redshift and $L^*$ is the break luminosity of the AGN
luminosity function (\ref{eq:boyle}).}
\label{fig:bh_biasst}
\end{figure}

To illustrate the application of this result, we adopt the quasar luminosity
function from \citet{boy00},
\begin{equation}
  \Psi(\lambda)={\Psi_\ast\over 10^{-(1+\alpha)(\lambda-\lambda^*)} + 
  10^{-(1+\beta)(\lambda-\lambda^*)}},
\label{eq:boyle}
\end{equation}
with $\alpha=-3.4$ and $\beta=-1.6$, where $\lambda^*$ is the ``break''
luminosity. The corresponding bias $\Delta\log_{10}M_\bullet$ is shown as a
function of the lower luminosity limit $\lambda_0-\lambda^*$ in Figure
\ref{fig:bh_biasst}.  Note that the bias in Figure \ref{fig:bh_biasst} becomes
smaller but does not vanish as the survey goes to fainter and fainter luminosity
limits: asymptotically, $\Delta\log_{10}M_\bullet\to-(1+\beta)\ln 10$ which
equals $0.12(\sigma_\mu/0.3)^2$.

\subsection{Bias in a Flux-Limited Survey}
\label{sec:bhmasstwo}

As a second example, we examine the bias in a flux-limited sample of AGN.  We
assume that the survey galaxies are distributed uniformly in Euclidean space,
that the probability distribution of AGN luminosities is given by equation
(\ref{eq:probone}), and that the survey contains all galaxies with flux
$f=L_{\rm AGN}/r^2$ exceeding some limiting flux $f_0$.

The bias is then given by
\begin{equation}
\Delta\log_{10}M_\bullet={\int d\lambda 
\int g(s)nV(\lambda)\,ds\int (\mu-a-bs)h(\mu-a-bs)\psi(\lambda-\mu)\,d\mu\over 
\int d\lambda 
\int g(s)nV(\lambda)\,ds\int h(\mu-a-bs)\psi(\lambda-\mu)\,d\mu},
\end{equation}
where $V(\lambda)=\frac13\Delta\Omega (L_{\rm AGN}/f_0)^{3/2}\propto
10^{3\lambda/2}$ is the volume within which an AGN of luminosity $\lambda$ can
be detected, $\Delta\Omega$ is the solid angle covered by the survey, and $n$
is the number density of galaxies. We have
\begin{eqnarray}
\Delta\log_{10}M_\bullet&=& {\int d\lambda \int
g(s)10^{3\lambda/2} \,ds\int (\mu-a-bs)h(\mu-a-bs)\psi(\lambda-\mu)\,d\mu\over
\int d\lambda \int g(s)10^{3\lambda/2}\,ds\int
h(\mu-a-bs)\psi(\lambda-\mu)\,d\mu} \\ &=& {\int 10^{3\lambda/2}\,d\lambda 
\int g(s)\,ds\int x\,h(x)\psi(\lambda-x-a-bs)\,dx\over \int
10^{3\lambda/2}\,d\lambda \int g(s)\int h(x)\psi(\lambda-x-a-bs)\,dx} \\ &=&
{\int 10^{3\lambda/2}\,d\lambda \int x\,h(x)\phi(\lambda-x)\,dx\over
\int 10^{3\lambda/2}\,d\lambda \int h(x)\phi(\lambda-x)\,dx},
\label{eq:xxx}
\end{eqnarray}
where $\phi(y)$ is defined by equation (\ref{eq:defphi}). To evaluate the
inner integral in the numerator of equation (\ref{eq:xxx}) we use equation
(\ref{eq:parts}), which yields 
\begin{eqnarray}
\Delta\log_{10}M_\bullet&=&-\sigma_\mu^2
{\int 10^{3\lambda/2}\,d\lambda {d\over d\lambda}\int
h(x)\phi(\lambda-x) \,dx\over
\int 10^{3\lambda/2}\,d\lambda  \int h(x)\phi(\lambda-x)\,dx}\\
&=&{3\ln 10\over 2}\sigma_\mu^2,
\label{eq:fluxlim}
\end{eqnarray}
where the last line follows from an integration of the numerator by parts.

\subsection{Selection Effects That Depend Only on Galaxy Properties}
\label{sec:prop}

The results of the previous subsections depend on the assumption that the
probability distribution of AGN luminosity is determined by the black-hole
mass and not the galaxy properties (eq.~\ref{eq:probone}). A foil to this is
to assume that the luminosity is determined by the galaxy properties and not
the black-hole mass. Thus equation (\ref{eq:probone}) is replaced by
\begin{equation}
  dp=\chi(\lambda-ks)\,d\lambda\quad\hbox{where}\quad \int\chi(x)dx=1  
\label{eq:probtwo}
\end{equation}
where $k$ is a constant. In this case $\Delta\log_{10}M_\bullet=0$ in any
flux-limited or luminosity-limited sample \citep{sal07}. 

\section{Correcting for Selection Bias May be Extremely Difficult} 

\label{sec:difficult}

Given the example bias calculations shown in the previous section, it may be
tempting to conclude that similar bias corrections may be estimated and
applied after the fact to existing surveys to determine the 
$M_\bullet-\sigma$ or $M_\bullet-L$ relations. We believe that in general such
corrections are difficult or impossible to apply reliably to current surveys,
for the following reasons.

\subsection{The Error Model is Poorly Known}

In the examples above, we assumed that at any galaxy property $s,$
$\mu=\log_{10}M_\bullet$ followed a normal distribution about the mean
$M_\bullet-s$ relation $\mu=f(s)$, with standard deviation characterized by a
cosmic scatter $\sigma_\mu.$ We saw that the bias is very sensitive to
$\sigma_\mu$ (in most examples $\propto\sigma_\mu^2$) yet this parameter is
poorly known for either the $M_\bullet-\sigma$ or $M_\bullet-L$ relations, in
part because an accurate estimate of $\sigma_\mu$ requires knowing the
observational errors in the $M_\bullet$ determinations. \citet{novak} have
studied this problem and conclude only that $\sigma_\mu\lesssim0.3$ for the
$M_\bullet-\sigma$ relation and $\sigma_\mu\lesssim0.5$ for the $M_\bullet-L$
relation.  Moreover, the \citet{novak} analysis assumes, as we do, that
$\sigma_\mu$ is constant over the parameter range of the $M_\bullet$
relations; the sample of galaxies with well-determined $M_\bullet$ is simply
too small to explore the possibility that it is not.

Even if an accurate estimate of $\sigma_\mu$ were available, the functional
form of the distribution of $\mu-f(s)$ is unknown.  The assumption of a normal
distribution is an obvious first step, yet the steep falloff of the $L$ or
$\sigma$ distribution at large values means that the selection bias for large
black-hole masses is sensitive to the wings of this distribution, where
$\mu-f(s)\gtrsim 2\sigma_\mu$. The assumed normal distribution is likely to
{\it under\,}estimate the selection bias since more realistic distributions
have fatter tails. Given that the local black-hole sample is so small that
$\sigma_\mu$ is poorly determined, the sample of galaxies with good
$M_\bullet$ determinations would have to be several orders of magnitudes
larger before an accurate form for the scatter function could be determined.

\subsection{The $M_\bullet$ Relationships are Poorly Known}

The $M_\bullet$ relations are imperfectly known at $z=0,$ and in particular
the range of $L$ or $\sigma$ that is used to determine them is rather limited.
For example, in the sample of 31 local galaxies with measured black-hole
masses used by \cite{tr02} to determine the $M_\bullet-\sigma$ relation, the
interquartile ranges in $L$ and $\sigma$ are only factors of 5.3 and 1.6
respectively. Thus, the calculation of the selection biases for
$M_\bullet\gtrsim 10^9M_\odot$ must contend with the fact that there are only
a few black holes observed in this mass range in the local sample (4 in the
\citealt{tr02} sample).
\citet{l07} show that the $L-\sigma$ relation must be
curved (in logarithmic coordinates) in the sense that $\sigma$ appears to
increase only slowly (if at all) with $L$ for the most luminous galaxies.
This implies that the present log-linear $M_\bullet-\sigma$ and $M_\bullet-L$
relation cannot be consistent at high galaxy luminosity.  The joint
$M_\bullet-s$ distributions shown in Figure \ref{fig:2df} are thus based on
extrapolated estimates at the highest $M_\bullet$ values that are likely to
change if more and better determinations of black-hole masses
$M_\bullet\gtrsim 10^9M_\odot$ become available.

\subsection{What Determines AGN Luminosity?}

The heart of our analysis of selection bias in \S\S\ref{sec:bhmass} and
\ref{sec:bhmasstwo} is the assumption that the black hole determines the AGN
luminosity, or more precisely that the galaxy properties do not.  This is
clearly true if, for example, (i) black holes radiate either at the Eddington
luminosity $L_{\rm Edd}$ or not at all, and (ii) the bolometric correction is
independent of galaxy properties. This assumption is also true in more general
circumstances, for example, (i) can be replaced by the weaker assumption that
the probability that a black hole is radiating at some $L_{\rm AGN}$ depends
only on the ratio $L_{\rm AGN}/L_{\rm Edd}$ (eq.\ \ref{eq:probone}).  This
assumption may be approximately correct: for example, \cite{hop06} find that
their simulations are well-fit by a model in which the probability
distribution of AGN luminosities depends only on $L_{\rm AGN}/L_{\rm peak}$
where $L_{\rm peak}$ is the peak luminosity of the AGN and $L_{\rm
peak}\propto L_{\rm Edd}^{1.12}$. Nevertheless, if galaxy properties do affect
the AGN luminosity---most likely by determining the feeding rate of matter
onto the black hole at luminosities $L\ll L_{\rm Edd}$---then the biases will
be different from those presented in \S\S\ref{sec:bhmass} and
\ref{sec:bhmasstwo}. In the limiting case that the AGN luminosity is
determined entirely by galaxy properties rather than black-hole mass, as might
be plausible for low-luminosity AGNs, there is no selection bias in a
flux- or luminosity-limited survey (\S\ref{sec:prop}).

\subsection{Did the Galaxy Make the Black Hole, or the Black Hole the Galaxy?}

We have assumed that the joint probability distribution of black-hole mass
$\mu$ and galaxy property $s$ can be modeled by equation (\ref{eqn:joint}).
This corresponds to a physical model in which the galaxy property is the
independent variable, and the black-hole mass is determined by the properties
of the galaxy through the cosmic-scatter function $h(y)$ and the $M_\bullet$
relation $f(s)$. This assumption is convenient for modeling the local sample
of galaxies with $M_\bullet$ determinations, which was selected by galaxy
properties, plus the ready knowledge of $g(s),$ the volume distribution of
$s.$ The alternative model provided by equation (\ref{eqn:jointa}), in which
the black-hole mass is the independent variable and the galaxy property is
determined by the black hole through the function $p(y)$, is much more
difficult to fit to the local observations as we have little direct knowledge
of $\Phi_\bullet(\mu),$ the black-hole mass function. However, our present
understanding of black-hole and galaxy formation does not allow us to say
which model, if either, is correct.

\subsection{Survey Selection Effects May be Poorly Known}

Most of our estimates of selection bias have been based on the assumption that
surveys are complete to a given AGN luminosity or flux, but this is an
oversimplification. Any survey that is based on measurements of the luminosity
of the host galaxy requires that the host is bright enough and large enough to
be separated from the AGN. Any survey that uses the velocity dispersion
measured from stellar absorption lines requires that the absorption lines are
strong enough, and that AGN emission lines in the vicinity of the absorption
lines are weak enough, to allow a reliable dispersion measurement. At the
opposite extreme, surveys of low-luminosity AGNs require that the emission
lines are strong enough to be detected against the continuum flux from the
galaxy. In such studies the selection depends in a complex way on properties
such as the ratio of AGN to galaxy luminosity, and on galaxy properties other
than $L$ or $\sigma$, such as the effective radius or central surface
brightness.

\section{Bias in Existing Surveys}

We have intended this paper mainly as a planning guide for future surveys to
explore the evolution of the $M_\bullet$ relation, rather than as a critique
of existing surveys. The following brief comments on existing surveys are
mostly intended to illustrate the application and impact of the estimates of
selection bias that we have made in earlier sections.

Several authors have determined the $M_\bullet-\sigma$ relation for nearby
AGNs, and compared this to the local relation for inactive galaxies. Using
seven Seyfert 1 galaxies with black-hole masses measured by reverberation
mapping, \cite{geb00} find $\Delta\log_{10}M_\bullet=-0.21\pm0.13$. For 16
AGNs with reverberation-based black-hole masses \cite{onk04} find
$\Delta\log_{10}M_\bullet=-0.26\pm0.15$ (although Onken et al.\ interpret
their result in terms of calibrating the reverberation-mapping method rather
than an offset in the $M_\bullet-\sigma$ relation). \citet{bar05} have
compared the black-hole masses in dwarf Seyfert 1 galaxies to an extrapolation
of the $M_\bullet-\sigma$ relation to lower dispersions; using black-hole
masses based on the photoionization method calibrated by the standard
``isotropic'' formula, they find $\Delta\log_{10}M_\bullet=-0.04$
(although they interpret their results with Onken et al.'s calibration and
thus find $\Delta\log_{10}M_\bullet=+0.23$).  \citet{greene06} have measured
the $M_\bullet-\sigma$ relation in 88 nearby AGN using the photoionization
method to determine $M_\bullet$---the sample is much larger than that of
\citet{onk04} but the photoionization mass estimates are less direct than those
from reverberation mapping. They find $\Delta\log_{10}M_\bullet=-0.21\pm0.06$.

Similarly, \cite{lab06} have determined the $M_\bullet-L$ relation from a
sample of 29 quasars with $z<0.6$; they determine black-hole masses using the
photoionization method and luminosities using HST photometry. They find
$\Delta\log_{10}M_\bullet=-0.36$, although they interpret their result in
terms of recalibrating the photoionization method. 

A common feature of the \cite{geb00}, \cite{onk04}, \citet{bar05}, and
\citet{greene06} samples is that their black-hole masses are relatively
modest: generally $M_\bullet<10^8M_\odot$ and often more than an order of
magnitude smaller.  Selection bias occurs at all $M_\bullet,$ depending on the
local slopes of the $L$ and $\sigma$ distributions, but these samples are
safely removed from the very strong biases seen at large values of $L$,
$\sigma$, and $M_\bullet$, where the distribution functions are falling
rapidly. Although we have emphasized selection bias in the context of
low- and high-redshift samples with similar black-hole masses, one selected by
galaxy properties and one by black-hole mass, selection bias can also arise if
the two samples are selected in the same way but are centered on different
black-hole mass ranges. 

We further note that selection effects in these samples are difficult to
model---for example, (i) as \cite{geb00} point out, the AGN variability
timescale depends on luminosity, and if the timescale is too long,
reverberation mapping is impractical; (ii) to be able to measure the velocity
dispersion requires some minimum ratio of the luminosity of the host galaxy to
the luminosity of the AGN. If we make the simplest possible assumption, that
the samples of nearby galaxies are complete but flux-limited, then equation
(\ref{eq:fluxlim}) implies $\Delta\log_{10}M_\bullet=0.31(\sigma_\mu/0.3)^2$.

\citet{woo06}, building on the initial work of \citet{tr04}, examine a sample
of 14 Seyfert 1 galaxies at $z=0.36\pm0.01$, measuring the velocity
dispersions from stellar absorption lines and the black-hole masses by the
photoionization method. Compared to the local sample of inactive galaxies with
measured black-hole masses, they find
$\Delta\log_{10}M_\bullet=0.62\pm0.10\pm0.25$; note that this result is based
on a calibration of the photoionization masses that {\it assumes} that the
\citet{onk04} sample should have exactly the same $M_\bullet-\sigma$ relation
as inactive galaxies, which need not be the case if selection bias is
accounted for. \cite{woo06} argue that their offset $\Delta\log_{10}M_\bullet$
is robust because the black-hole masses are measured by the same technique,
with the same calibration, in the local and high-redshift samples, but the
selection bias in the two samples is likely to be quite different; the
\citet{onk04} sample is more nearly flux-limited, while the \citet{woo06}
sample is more nearly luminosity-limited, and the selection bias in these two
cases is quite different (eqs.\ \ref{eq:fluxlim} and \ref{eq:biasone}).  A
useful next step in assessing the bias would be to determine the luminosity
function for AGN selected by the criteria used by \cite{woo06}, for use in
equation (\ref{eq:biasone}).

It may also be significant that the typical
$M_\bullet\approx4\times10^7M_\odot$ in the \citet{onk04} sample is roughly an
order of magnitude less massive than the typical
$M_\bullet\approx3\times10^8M_\odot$ in the \citet{woo06} sample.  As is
evident in Figure \ref{fig:bh_bias}, the bias is a strong function of
$M_\bullet,$ and indeed the bias in the $M_\bullet-\sigma$ relation changes
sign at $M_\bullet\sim10^8M_\odot$.  Thus we would expect that selection bias
might affect the results even if the \citet{onk04} sample were chosen in
precisely the same way as the \citet{woo06} sample. 

Figure \ref{fig:bh_bias} predicts a bias of $\Delta\log_{10}M_\bullet\sim0.2$
at $M_\bullet\approx3\times10^8M_\odot$ for $\sigma_\mu=0.30$. This is only
40\% of the offset claimed by \citet{woo06}. Thus, it appears that selection
bias cannot explain all or even most of the \citet{woo06} $\Delta
\log_{10}M_\bullet$; however, given the uncertainties discussed in the
previous section, the possibility that the correction may be as large as the
claimed offset cannot be ruled out.

\citet{peng} find $\Delta\log_{10}M_\bullet\simeq -0.1$ compared to the local
$M_\bullet-L$ relation, using a sample of 11 quasars at $z\simeq 2$; the
black-hole masses are determined with the photoionization method. If we assume
that the velocity dispersion of the host galaxy does not change between
$z\simeq2$ and $z=0$, and that the luminosity of the host fades by 1--2
magnitudes as predicted by passive stellar evolution, then this result
corresponds to $\Delta\log_{10}M_\bullet\simeq 0.4$--0.8 as measured by the
$M_\bullet-\sigma$ relation. However, the potential selection bias is rather
large for this sample, partly because the cosmic scatter in the $M_\bullet-L$
relation may be larger than in the $M_\bullet-\sigma$ relation. If the sample
is luminosity limited, the model shown in the left panel of Figure
\ref{fig:bh_bias} predicts $\Delta\log_{10}M_\bullet=(0.3\hbox{--}0.7)$ for
$\sigma_\mu=0.5$ for the range of black-hole masses in the sample, and the
flux-limited model predicts $\Delta\log_{10}M_\bullet=0.58$ for
$\sigma_\mu=0.5$ (note that our assumption of a Euclidean metric is not valid
at $z\simeq2$, and should be generalized to a realistic cosmological model). A
further complication is that their method requires measuring the bulge
luminosity in the presence of the bright AGN, so imposes a selection on the
ratio of AGN to galaxy luminosity.
Without an accurate estimate of the selection bias for the \citet{peng} sample,
their measurement of $\Delta \log_{10} M_\bullet$ does not provide
clear evidence for rapid evolution in the ratio of black-hole mass to
stellar mass in the galaxy. 

\citet{sal07} find that galaxies of a given dispersion at $z\simeq 1$ have
black-hole masses that are larger by $\Delta\log_{10}M_\bullet\sim 0.45$ than
at $z\simeq0$. In contrast to most other studies, they carefully consider the
effects of selection bias using Monte-Carlo models and conclude that for
$\sigma_\mu=0.3$ selection bias increases $\log_{10}M_\bullet$ by $\sim 0.1$
at low redshift and $0.2$ at $z\simeq 1$, contributing a net offset of
$0.1$. They also argue that scatter in the relationship between the width of
the [O III] emission line and the true stellar dispersion contributes an
additional offset of $\sim 0.15$, so after correcting for these biases
$\Delta\log_{10}M_\bullet\sim 0.2$. Their estimates of selection bias are
similar to, but somewhat smaller than, the bias that we estimate for their
sample from equations (\ref{eq:biasone}) and (\ref{eq:fluxlim}), assuming that
the low-redshift sample is flux-limited and the high-redshift sample is
luminosity-limited; we believe our estimates are more robust because
they do not depend on an assumed form for the distribution $\psi(\lambda-\mu)$
of AGN luminosities at a given black-hole mass.

Finally, we note that theoretical models of the evolution of the
$M_\bullet-\sigma$ relation \citep{rob06} predict
$\Delta\log_{10}M_\bullet=-0.18$ at $z=2$. Strong selection bias could reduce
or eliminate the difference between this prediction and the positive values of
$\Delta\log_{10}M_\bullet$ found at high redshift in most of the above
studies. 

\section{Recapitulation} 

We have described a selection bias that affects investigations into the
cosmological evolution of the $M_\bullet-\sigma$ or $M_\bullet-L$ relations.
We summarize the main points as follows:

\begin{itemize}

\item Cosmic scatter in the $M_\bullet-\sigma$ or $M_\bullet-L$ relations
  means that a galaxy of given $L$ or $\sigma$ will host black holes with a
  range of $M_\bullet$ values.

\item The steep decline in the luminosity function for the most luminous
  galaxies means that the rare occurrence of high-mass black holes in numerous
  ``modest'' galaxies overwhelms the frequent occurrence of black holes of
  similar mass in the rare galaxies with very high luminosity. A similar
  conclusion applies to dispersion instead of luminosity.

\item Selection of a sample of black holes hosted by inactive galaxies at low
  redshift explores the distribution of black-hole masses $M_\bullet$ that a
  galaxy will host, given its luminosity $L$ or dispersion $\sigma.$ Selection
  of a sample by AGN luminosity at high redshift explores the distribution of
  $L$ or $\sigma$ at a given $M_\bullet$ (if, as is likely, the distribution
  of AGN luminosity is determined mainly by the black-hole mass rather than
  the galaxy properties). Confusing the two distributions may create a false
  signature of evolution.

\item The selection bias is substantial for the current estimates of the
  cosmic scatter in the $M_\bullet-L$ or $M_\bullet-\sigma$ relations,
  particularly for galaxies containing the most massive black holes. Neither
  the root-mean-square cosmic scatter nor the shape of the scatter function is
  known accurately and hence the bias cannot be reliably corrected for, given
  our present and even foreseeable state of knowledge.

\item Our rough estimates of selection bias as applied to a number of
  evolutionary studies show that the potential bias may be larger than many of
  the $\Delta \log_{10}M_\bullet$ or $\Delta s$ values observed.  With the
  notable exceptions of \citet{fi06} and \citet{sal07}, it appears that most
  studies of the evolution of the $M_\bullet$ relations have not
  considered the effects of selection bias.  In most cases, accounting
  properly for selection bias is likely to reduce the the evolution in the
  $M_\bullet$ relations that has been observed in recent surveys.

\item The only way to avoid selection bias is to choose high- and low-redshift
  galaxy samples using precisely defined, objective criteria that are
  precisely the same for the two samples. To our knowledge, this has not yet
  been done.

\end{itemize}

\acknowledgments

The work was initiated during a Lorentz Center workshop attended by the
authors.  We thank the staff of the Lorentz Center and Dr. Tim de Zeeuw for
hosting us.  We thank Drs. Karl Gebhardt, Richard Green, and Nadia Zakamska
for useful conversations.

\clearpage

\end{document}